\documentclass[12pt]{article}
\usepackage{amsfonts}
\usepackage{amssymb}
\usepackage{amsmath,amsfonts}
\usepackage{graphicx}
\makeatletter\@addtoreset{equation}{section}\makeatother

\newcommand{\be}{\begin{equation}}
\newcommand{\ee}{\end{equation}}
\newcommand{\bea}{\begin{eqnarray}}
\newcommand{\eea}{\end{eqnarray}}

\newcommand{\ve}{\epsilon}

\def\bsp{\be\begin{split}}

\def\p{\partial}

\def\SO{\mathrm{SO}}

\def\SU{\mathrm{SU}}

\def\inc#1#2{{%
\raise-.5\ht1\box1%
}}

\newcommand{\Tr}{{\rm Tr\,}}
\newcommand{\tr}{{\rm tr\,}}

\newcommand{\cN}{{\mathcal N}}

\renewcommand{\title}[1]{\vbox{\center\LARGE{#1}}\vspace{5mm}}
\renewcommand{\author}[1]{\vbox{\center#1}\vspace{5mm}}
\newcommand{\address}[1]{\vbox{\center\em#1}}
\newcommand{\email}[1]{\vbox{\center\tt#1}\vspace{5mm}}

\begin{document}
\begin{titlepage}
\begin{center}
\vspace{5mm}
\hfill {\tt }\\
\vspace{20mm}
\title{Wilson Loop in $\cN=2$ Quiver/M theory Graivty Duality}
\vspace{10mm}
\author{\large Yang Zhou}

\address{Institute of Theoretical Physics\\
Chinese Academic of Science, Beijing 100190, PRC}

\email{yzhou@itp.ac.cn}

\end{center}
\vspace{10mm}

\abstract{
\noindent
We study Wilson loops in the 4-dimensional $\cN=2$ supersymmetric
quiver/M theory duality recently constructed by Gaiotto and Maldacena, that is conjectured to be dual to M2 branes on $AdS_5\times S^4$ fibered over $\Sigma_2$.
 We use the localization method raised by Pestun, in order to compute the  Wilson loop in 4-dimensional $\cN=2$ supersymmetric linear quiver gauge theory. We obtained the consistent results in the quiver gauge theory and the 11-dimensional supergravity. This match supports the $\cN=2$ quiver SCFT / M theory supergravity duality.
}

\vfill

\end{titlepage}


\section{Introduction}

We study supersymmetric Wilson loop operators in the
four-dimensional quiver theory of Gaiotto and Maldacena
~\cite{Gaiotto:2009gz}. This
theory is conjectured to represent the low-energy dynamics of $N$
coincident M5-branes wrapping on a Riemann surface. This in turn has an alternative description
as one M theory on $AdS_5\times S^4$ fibered over $\Sigma_2$ at IR fixed point, where $\Sigma_2$ is a Riemann surface with constant curvature. Though there is a beautiful pre-map between $\cN=2$ quiver/M theory duality~\cite{Gaiotto:2009gz}, there remains many details which need to match between field theory and M theory, like supergravity modes from Kaluza Klein reduction and spectrum of quiver fields. One typical question is the duality between Wilson loop operators in the quiver and corresponding membranes in the bulk, which is investigated in the present work.

We shall focus on Wilson loop operators due to well known reasons. One is that they are the principal gauge invariant observables.  Another, these operators play an
important role in the gauge/gravity correspondence~\cite{Maldacena:1997re},
since they are found to be dual to semiclassical
strings in the dual supergravity background~\cite{Rey,Maldacena-wl}. Third, it can be calculated exactly in some gauge theories like $\cN$=4 SYM even at strongly coupled case and
matched with string theory~\cite{Erickson:2000af,Drukker:2000rr,Pestun:2007rz}.
It is therefore interesting to see if an
analog observable exists in the new $\cN$=2 4D quiver gauge theory which is expected to describe the IR field theory of M5 branes wrapped over Riemann surface.

\section{Wilson loop observable in Gauge theory}

We shall first introduce the quiver/M theory duality. Recently, a method to construct a large kind
of 4D $\cN=2$ superconformal field theories (SCFTs)
was discovered in \cite{Gaiotto:2009we} by Gaiotto.
By analyzing Seiberg-Witten curves for
quiver theories based on  $\SU$ gauge groups \cite{Witten:1997sc},
Gaiotto gives a kind of SCFTs which arises as a compactification on a Riemann surface with punctures of the six-dimensional $A_{N-1}$ theory with $(2,0)$ supersymmetry.
The marginal couplings of a quiver theory are encoded in the moduli of the Riemann surface, and both weakly-coupled and
strongly-coupled limits were shown to correspond to degenerations
of the Riemann surface.
Gaiotto's approach shows us a new unified understanding of the S-dualities of
$\SU(2)$ gauge theory with four flavors \cite{Seiberg:1994aj},
which involves the triality of $\SO(8)$ flavor symmetry,
and of $\SU(3)$ gauge theory with six flavors,
the strongly-coupled limit of which is conjectured to be dual to the mysterious isolated SCFT with $E_6$ flavor symmetry \cite{Minahan:1996fg} with a gauged $\SU(2)$ part and one flavor \cite{Argyres:2007cn}.
More important, it predicted a totaly new family of SCFTs $T_N$, with $\SU(N)^3$ flavor symmetry which are isolated, with no marginal couplings. Gaiotto duality has been extended in many directions, like D type~\cite{Tachikawa:2009rb}, different matter~\cite{Nanopoulos:2009xe}, more complex brane webs~\cite{Benini:2009gi}. Due to the clear framework, in which 4D $\cN=2$ quiver gauge theories are realized by 6D (2,0) $A_{N-1}$ SCFT over Riemann surfaces, a new connection between Nekrasov partitions functions 4D field theory and Liouville correlation functions has been proposed~\cite{Alday:2009aq}, and there are related extensions~\cite{NLC}. In the interesting work \cite{Gaiotto:2009gz},
quivers constructed with the theory $T_N$ were considered as the IR field of the $N$ M5 branes wrapped over Riemann surface with punctures, investigated more concretely. In particular, the holographic gravity solutions were discussed in detail and various quantities like central charges of conformal anomaly and dimensions of a certain kind of special operators were matched between the field theory and M-theory description. In the following subsection, we will first understand this duality in details and study how to construct Wilson loop operators in the next subsection.

\subsection{Superconformal quivers}

We begin with a quiver gauge theory in 4D first considered in \cite{Witten:1997sc}. In 10D IIA theory, considering intersecting N D4 branes and NS5 branes of the following configuration:
 \bea\begin{split}
 NS5:&1\quad 2\quad 3\quad 4\quad 5\quad \ast\quad \ast\quad \ast\quad \ast\\
 D4: &1\quad 2\quad 3\quad \ast\quad \ast\quad 6\quad \ast\quad \ast\quad \ast
 \end{split}
 \eea
 N D4 branes are divided into $N+1$ parts by $N$ parallel five branes in the $x^4$ direction, then we obtain a macroscopically low energy 4D conformal qiver gauge theory with $N-1$ gauge groups $\prod SU(N)$, $N$ fundamental hypermultiplets at each of the two ends and bi-fundamental hypermultiplets between each two gauge groups.
 We refer to the quiver diagram in Fig.2 (a) in \cite{Gaiotto:2009gz}. Similar to Argyres-Seiberg duality \cite{Argyres:2007cn}, the strongly coupled
 limit of the original quiver produce alternative weakly coupled dual gauge groups, with a piece of strongly coupled $T_N$, which is presented by the
  dual quiver in Fig.2(b) in \cite{Gaiotto:2009gz}. $T_N$ is $\cN=2$ SCFT with $SU(N)^3$ global symmetries and no coupling constant, clear in Fig.2(c) in \cite{Gaiotto:2009gz} if
   we zero couple the $SU(N)$ in Fig.2(b) in \cite{Gaiotto:2009gz}.

Before constructing wilson loop operators, we list the already known information about the mysterious $T_N$.

1. The contribution to the $SU(N)$ beta function from the $T_N$ part in Fig.2(b) in \cite{Gaiotto:2009gz} is the same as that of $N$ fundamental hypers, which makes the whole quiver is conformal.

2. An isolated $T_N$ describes the low energy field of $N$ M5 branes wrapped over a Riemann surface with three punctures, with three isolated $SU(N)$ global symmetries, no coupling constant.

3. The effective number of vectormultiplets and hypermultiplets in $T_N$ is $n_v$ and $n_h$
\be
n_v={2N^3\over 3}-{3N^2\over 2}-{N\over 6}+1, \quad\quad n_h={2N^3\over 3}-{2N\over 3},
\ee
and the conformal anomaly
\be
c={2n_v+n_h\over 12} \quad\quad a={5n_v+n_h\over 24}.
\ee
By Gaiotto duality techniques, we can construct more general quiver diagram by sewing $T_N$ by $SU(N)$ vectormultiplets. Thus for a genus $g$ quiver, we obtain effective vectors and hypers
$$n_v^g=(g-1)[4N^3/3-N/3-1]$$
$$n_h^g=(g-1)[4N^3/3-4N/3]\;.$$
Thus $a$ and $c$ for genus $g$ quiver are
\be
c^g=(g-1)[N^3/3-N/6-1/6]\;,
\ee
\be
a^g=(g-1)[N^3/3-N/8-5/24]\;.
\ee
\subsection{Wilson loop}
We shall study Wilson loop operators of the quivers in above subsection. As a warm up, in $\cN=4$ SYM, the Wilson loop is constructed by
\be
W(C)={1\over N}\Tr[P\exp(\oint(iA_\mu \dot{x}^\mu + \theta^I\phi^I\sqrt{\dot{x}^2})d\tau)],
\ee
where $x^\mu(\tau)$ parametrizes the loop and $\theta^I$ is the direction vector. The added adjoin scalar helps the Wilson operator keep some supersymmetries. To obtain the Wilson loop in quiver in Fig.2(a) in \cite{Gaiotto:2009gz}, we first consider the simplest two gauge group quiver
\be
W^{(1,2)}(C)={1\over N}\Tr[P\exp(\oint(iA_\mu^{(1,2)} \dot{x}^\mu + i\Phi_0^{(1,2)})d\tau].
\ee
Actually, if we have a closed quiver with $N$ gauge groups and $N$ bi-fundamentals, which is a closed form of quiver in Fig.2(a) in \cite{Gaiotto:2009gz}. We can construct the same supersymmetric Wilson loop for each gauge group $A^{(n)}$, $n=1\sim N$.
\be\label{loop1}
W^{(n)}(C)={1\over N}\Tr[P\exp(\oint(iA_\mu^{(n)} \dot{x}^\mu + i\Phi_0^{(n)})d\tau].
\ee
We can go straight forward to the perturbative calculation for the once given Lagrangian. For the 3D ABJM\footnote{Where, there is a adjoin scalar composed of two bi-fundamental scalars.}, perturbative calculation shows that Wilson loop on a circular~\cite{Chen:2008bp,Rey:2008bh,Drukker:2008zx}
\be
\langle W^{(1,2)}\rangle=1+{\pi^2N^2\over k^2}-{\pi^2N^2\over 6k^2}+O(k^{-3})\;.
\ee
In the strongly coupled range, it is expected that there exists a combined gauge invariant Wilson loop
\be\label{averageloop}
W_0={1\over N}(W^{(1)}+W^{(2)}+\cdots+W^{(n)}),
\ee
which is shown to be dual to the fundamental string / or M brane with two ends on the boundary loop in gauge/gravity duality. To support this expectation, we need to go to the matrix description of Wilson loop operators.
In superymmetric gauge theory side, one exact calculation by localization was done by Pestun~\cite{Pestun:2007rz}, which proved the Gaussian matrix conjecture of~\cite{Erickson:2000af,Drukker:2000rr} in $\cN=4$ SYM and gave the $\cN=2$ method in the same time. This localization method, first introduced in~\cite{Nekrasov:2002qd} will be discussed in the following section more concretely, and we will use the main result for the Wilson loop in $\cN=2$ quivers here at first and leave the provement in the following section. As a warm up, in $\cN=4$ SYM, the expectation value of a Wilson loop on a circular
\be
\langle W(C)\rangle=\langle {1\over N}\tr e^{2\pi\Phi}\rangle={2\over \sqrt{2\lambda}}I_1(\sqrt{2\lambda}).
\ee
In the large $N$ limit,
\be
{2\over \sqrt{2\lambda}}I_1(\sqrt{2\lambda})\simeq{2\over \pi}(2\lambda)^{-3/4}e^{\sqrt{2\lambda}}.
\ee
In field theory, this result comes from the summation of all ladder diagrams and the 1-loop correction is proved to be trivial. While in $\cN=2$ SYM, the 1-loop corrections nontrivial, thus the matrix model is somehow hard to solve. However, taking account of the 1-loop effect, in the $\cN=2$ quiver with $N$ $SU$ gauge field and $N$ bi-fundamentals, the combined operator has the same equation of motion as the $\cN=4$ Wilson operators. 
The final effective t'Hooft coupling for $W_0(C)$ is
\be\label{coupling}
 {1\over \lambda}={1\over N}({1\over \lambda_1}+{1\over \lambda_2}+\cdots+{1\over \lambda_N}).
\ee Thus, the result is that VEV of the ``average Wilson
loop"(\ref{averageloop}) is the same as $\cN=4$ SYM with a new
t'Hooft coupling~(\ref{coupling}). Let us turn to the quivers in Fig
2 in \cite{Gaiotto:2009gz}. In (a), by (\ref{coupling}) the VEV of
the Wilson loop  is \be \langle W_0\rangle_a={2\over
\sqrt{2\lambda_a}}I_1(\sqrt{2\lambda_a}). \ee And for the tail in
(b), the VEV of the Wilson loop is roughly\footnote{By ``roughly" we
mean, for quiver with different-rank gauge groups, the following
formula needs further proved. But quantitatively, $\langle
W_0\rangle_a$ dominates in the large $N$ limit, thus this
generalization does not effect the final result.} \be \langle
W_0\rangle_b={2\over \sqrt{2\lambda_b}}I_1(\sqrt{2\lambda_b}). \ee
Where \be \lambda_a={N^2\over g_1^{-2} +
g_2^{-2}+\cdots+g_N^{-2}},\quad \lambda_b={N\over g_2'^{-2}*{1\over
2}+\cdots+g_N'^{-2}*{1\over N}}\;. \ee For simplicity, we assume
$g_i$ goes to infinitely large in a certain way but keep
$g^{-2}=\sum g_i^{-2}$ fixed, and $g'=g'_2=\cdots=g'_N$ are weak
coupling constants. Then \be \lambda_a=g^2N^2,\quad
\lambda_b={g'^2N\over \ln N-1}. \ee At the large $N$ limit, there is
always $\lambda_a\gg\lambda_b$. We assume that there exists a term
we missed while constructing the Wilson loop operator in the whole
quiver (b), which comes from the $T_N$ theory. We assume it is a
adjoin term, which should be added into the integration in
(\ref{loop1}). In the decouple limit between $T_N$ and tail of (b),
we can define a loop operator in $T_N$ theory as\footnote{By
(\ref{fieldresult0}), concretely we define a loop operator in a
minimal block including one ``$T_N$'', which is half of the two
genus quiver, shown in Fig.5 a) in~\cite{Gaiotto:2009gz}.}
\be\label{fieldresult0} \langle W(C)\rangle_{T_N}={\langle
W(C)\rangle_a\over \langle W(C)\rangle_b}\simeq
\biggr({\lambda_a\over\lambda_b}\biggr)^{-3/4}e^{\sqrt{2\lambda_a}-\sqrt{2\lambda_b}}.
\ee In the large $N$ limit, the leading exponential term is
\be\label{fieldresult}
\langle W(C)\rangle_{T_N}\sim
e^{\sqrt{2}gN}.
\ee
This term will contribute mostly in the Wilson operators in elementary blocks of quivers composed of $T_N$s and other gauge fields. 
\subsection{Matrix model by localization}
In order to prove (\ref{coupling}), we need to take details of matrix conjecture for Wilson loop operators. First, let us briefly review the localization method to calculate the Wilson loop in $\cN=4$ SYM, which is presented in~\cite{Pestun:2007rz}. It is a complex process more or less, which starts  from the construction of $\cN=4$ SYM in S$^4$, we would like to list the main steps to give a simple description of the process. This can be extended directly to $\cN=2$ loop operators.

\underline{STEP 1.} Construction of $\cN=4$ SYM theory on S$^4$.
We start from the 10 dimensional $\cN=1$ SYM action
\begin{align}
  S =\int d^{10}x \frac {1} {2 g^2_{YM}} \left( \frac 1 2 F_{MN} F^{MN} - \Psi \Gamma^{M}
  D_{M} \Psi \right)\;,
\end{align}which
is invariant under the supersymmetry transformations
\begin{align*}
   \delta_{\epsilon} A_{M} = \epsilon \Gamma_{M} \Psi \quad\quad \delta_{\epsilon} \Psi = \frac 1 2 F_{MN} \Gamma^{MN} \epsilon.
\end{align*}
After dimension reduction, we have the $\cN=4$ SYM on S$^4$
\begin{equation}
  \label{eq:SYM_S^4_N=4}
  S_{\cN=4} = \frac 1 {2 g_{YM}^2} \int_{S^4} \sqrt{g} d^4 x \left(\frac 1 2 F_{MN} F^{MN}
    - \Psi \gamma^{M} D_{M} \Psi + \frac 2 {r^2} \Phi^A \Phi_A  \right),
\end{equation}
which is invariant under the $\cN=4$ super conformal transformations
\begin{align}
  \label{eq:SYM_S^4_N=4-trans}
   \delta_{\ve} A_{M} = \ve \Gamma_{M} \Psi \quad\quad
   \delta_{\ve} \Psi = \frac 1 2 F_{MN} \Gamma^{MN} \ve + \frac 1 2 \Gamma_{\mu A}
  \Phi^A \nabla^{\mu} \ve.
\end{align}

\underline{STEP 2.} Off-shell SUSY extention.
So far, we have superconformal algebra only closed on shell. To close off-shell SUSY of $\cN$=4 on S$^4$ we use the dimensional reduction of Berkovits method~\cite{Berkovits:1993zz}. We add 7 auxiliary fields $K_i$ with free quadratic action and the action becomes
\begin{equation}
  S_{\cN=4} = \frac 1 {2 g_{YM}^2} \int_{S^4} \sqrt{g} d^4 x \left(\frac 1 2 F_{MN} F^{MN}
    - \Psi \gamma^{M} D_{M} \Psi + \frac 2 {r^2} \Phi^A \Phi_A  \right)-K_iK^i,
\end{equation}
which is invariant under the superconformal transformations
\begin{equation}
  \begin{aligned}
    \label{eq:off-shell-susy}
    &  \delta_{\ve} A_{M} = \Psi \Gamma_{M} \ve \\
    &  \delta_{\ve} \Psi = \frac 1 2 \gamma^{MN} F_{MN} + \frac 1 2 \gamma^{\mu A} \phi_{A} D_{\mu} \ve + K^i \nu_i \\
    & \delta_{\ve} K_i = - \nu_i \gamma^{M} D_{M} \Psi,
  \end{aligned}
\end{equation}
where spinors $\nu_i$ with $i=1, \dots, 7$ are required to satisfy the condition in (2.29)-(2.31) in~\cite{Pestun:2007rz}.

\underline{STEP 3.} Localization by choosing $V$ potential.
We deform the action by $V$, which is a $Q$-exact term
\be
S\longrightarrow S+ tQV.
\ee
For $Q^2$-invariant $V$, the expectation value of Wilson loop does not change under the deformation. Thus, if we set $t\rightarrow\infty$, the action localizes to critical points of $QV$, over which we will integrate in the end. Choosing a suitable $V$, the action will become particular simple after localization.

Finally, the remaining nonvanishing action
\be
S[a]={4\pi^2r^2\over g^2}a^2\;,
\ee
where $a$ is an element of Lie algebra $G$. And the VEV of Wilson loop becomes a matrix model.
For $\cN=2$, the 1-loop contribution $Z^{\cN}_{\text{1-loop}}$ and instanton correction  $Z_{\textrm{inst}}^{\cN}$ is non-trivial, and the VEV of Wilson loop is given by
\begin{equation}
  \label{eq:main-result}
   Z^{\cN}_{S^4} \langle W_R(C) \rangle = \frac {1} {\textrm{vol}(G)} \int_G [da] \, e^{-\frac { 4 \pi^2 r^2} {g^{2}} (a,a) } Z^{\cN}_{\text{1-loop}}(ia)|Z_{\textrm{inst}}^{\cN}(r^{-1},r^{-1},ia)|^2 \tr_R e^{2\pi r i a}\;.
\end{equation}
In the present letter we do not focus on the instanton effect, but take account of 1-loop correction seriously.
\subsection{Computation}
First, we shall write the matrix model for $\cN=2$ $SU(N)$ Wilson loop specifically and give the final integration form. Then, we extend this matrix model to Wilson operators in $\cN=2$ quiver gauge theory. Finally, we show how to obtain quiver Wilson loop result (\ref{coupling}) form saddle point equations.
\subsubsection{$\cN=2$ SU(N) Wilson loop}
From Pestun's view, for all $\cN=2$ SCFT theory with massless hypermultiplets taken in representation by $W$. One can calculate the 1-loop contribution by
\begin{equation}
\label{eq:Z-1-loop-any-matter}
Z_{1-loop}^{\cN=2, W}(ia) = \frac{ \prod_{\alpha \in \text{weights}(\text{Ad})} H(i\alpha \cdot a/\ve) }
{ \prod_{w \in \text{weights}(W)} H(iw\cdot a/\ve) }\;.
\end{equation}
Where
\begin{equation}
   H(z) = e^{-(1+\gamma) z^2} \prod_{n=1}^{\infty} \left(1 - \frac {z^2}{n^2}\right)^{n} \prod_{n=1}^{\infty}
e^{\genfrac{}{}{}{1}{z^2} {n}}.
\end{equation}
The valid condition is
$$\sum_{\alpha} (\alpha \cdot a)^2 = \sum_{w} (w \cdot a)^2$$ for any $a$ in $G$. This condition implies the vanishing $\beta$-function for the $\cN=2$ theory with a hypermultiplet in $W$. The matrix model for the expectation value of the Wilson loop in the spin-$j$ representation is
\be
W=Z^{-1}\int_G da \, e^{-\frac { 4 \pi^2 r^2} {g^{2}} (a,a) }\frac{ \prod_{\alpha \in \text{weights}(\text{Ad})} H(i\alpha \cdot a/\ve) }
{ \prod_{w \in \text{weights}(W)} H(iw\cdot a/\ve) }\tr_R e^{2\pi r i a}.
\ee
Where $\ve=1/r$, and $r$ can be canceled by rescaling $a:=a\cdot r$. The integral can be switched to Cartan subalgebra. Then the above formula becomes
\be\label{wilsonF}
W=Z^{-1}\int_G da
 e^{-\frac { 4 \pi^2} {g^{2}} (a,a) }\frac{ \prod_{\alpha \in \text{weights}(\text{Ad})} H(i\alpha \cdot a) }
{ \prod_{w \in \text{weights}(W)} H(iw\cdot a) }\tr_R e^{2\pi  i a}.
\ee
Take the $\cN=2$ theory with the $SU(N)$ gauge group and 2N hypermultiplets in the fundamental representation.
In fundamental $R$, we have $N$ weights. In order to represent them, we choose one set of convenient basis for the fundamental representation
\be
V_i=(\cdots, 1_{(i)}, \cdots)^{\textrm{T}}, \quad i=1\sim N,
\ee
$i$ means the $i$th position. In the same time, we choose a set of basis for the $N\times N$ traceless Hermitean matrices, which represent the $N^2-1$ generators of $SU(N)$ as follow
$$
(T^{(1)}_{ab})_{cd}={1\over 2}(\delta_{ac}\delta_{bd}+\delta_{bc}\delta_{ad});\quad
(T^{(2)}_{ab})_{cd}={-i\over 2}(\delta_{ac}\delta_{bd}-\delta_{bc}\delta_{ad});\quad
(a<b;1\leq a,b\leq N)
$$
$$(T^{(3)}_a)_{cd}=
\delta_{cd}[2a(a-1)]^{-1/2} ~\textrm{if}~ c<a,
-\delta_{cd}[(a-1)/2a]^{1/2} ~\textrm{if}~  c=a,
0 ~\textrm{if}~  c>a. \quad(2\leq a\leq N)$$
$N-1$~~$T^{(3)}_a$ are Cartan generators, and we obtain the weight vector for the fundamental representation by acting $T^{(3)}$ on $V_i$. In adjoint representation, we have $N^2-1$ weights, which contains $N-1$ zero weights and $N(N-1)$ finite weights. These finite weights can be obtained by subtracting each two fundamental weights, whose number is just $2\times C^2_N=N(N-1)$. Thus, we can get $\alpha$ and $w$ for adjoint and fundamental representation respectively.
The diagonalized generators are
$$
T^{(3)}_a=diag(1, \cdots, 1, -(a-1), 0, \cdots, 0)/\sqrt{2a(a-1)}\;.
$$
The fundamental weights:
$$w_1=({1\over 2}, {1\over 2\sqrt{3}}, {1\over 2\sqrt{6}},\cdots, {1\over \sqrt{2a(a-1)}}, \cdots)\;,$$
$$w_2=(-{1\over 2}, {1\over 2\sqrt{3}}, {1\over 2\sqrt{6}},\cdots, {1\over \sqrt{2a(a-1)}}, \cdots)\;,$$
$$w_3=(0, {-2\over 2\sqrt{3}}, {1\over 2\sqrt{6}},\cdots, {1\over \sqrt{2a(a-1)}}, \cdots)\;,$$
$$w_4=(0~~, 0~~, {-3\over 2\sqrt{6}},\cdots, {1\over \sqrt{2a(a-1)}}, \cdots)\;,$$
$$\cdots$$
$$w_{N}=(0~~, 0~~, 0~~, \cdots, 0~~, {-(N-1)\over \sqrt{2N(N-1)}}).$$
The finite adjoint weight is
$$\{\alpha_1, \alpha_2, \cdots \alpha_{N(N-1)}\}=w_{12}, w_{21}, w_{13}, w_{31}, w_{23}, w_{32} \cdots, w_{N-1N}, w_{NN-1}\;.$$
where
$$w_{ij}=w_i-w_j.$$
More simply, we can take the Cartan as the set of diagonal matrices $a=diag(\lambda_1,\lambda_2,...\lambda_{N-1}, \lambda_N)$, with a zero summation condition. The roots of $G=SU(N)$ are labeled by integers $i\neq j$, and we have:
$$
\prod_\alpha \alpha\cdot a\sim\prod_{i\neq j}\alpha'_{ij},\quad\quad\alpha'_{ij}=\lambda_i-\lambda_j.
$$
Then (\ref{wilsonF}) becomes
\be
W=Z^{-1}\int \prod_{i< j} (\lambda_i-\lambda_j)^2 \, (\prod_i^{N-1} e^{-\frac { 4 \pi^2} {g^{2}} \lambda_i^2} d\lambda_i)\frac{ \prod_\alpha  H(i\alpha \cdot a) }
{ \prod_w H(iw\cdot a) }\tr_R e^{2\pi  i a}.
\ee
In fundamental representation of Wilson loop, the above formula gives
\be\label{loopintegrate}
W=Z^{-1}\int \prod_{i< j} (\lambda_i-\lambda_j)^2 \, (\prod_i^{N-1} e^{-\frac { 4 \pi^2} {g^{2}} \lambda_i^2} d\lambda_i)\frac{ \prod_{i\neq j} H(i\lambda_i-i\lambda_j) }
{ \big[\prod_i H(i\lambda_i)\big]^{2N} }\sum_i^{N}e^{2\pi \lambda_i}.
\ee
Remember
\begin{equation}
   H(z) = e^{-(1+\gamma) z^2} \prod_{n=1}^{\infty} (1 - \frac {z^2}{n^2})^{n} \prod_{n=1}^{\infty}
e^{\genfrac{}{}{}{1}{z^2} {n}}=G(1+z)G(1-z).
\end{equation}
For large $\lambda_i$, we expand
\begin{equation}
  \log G(1+z) = \frac 1 {12} - \log A + \frac {z} 2 \log 2 \pi + ( \frac {z^2} 2 - \frac 1 {12} ) \log z
  - \frac 3 4 {z^2} + \sum_{k=1}^{\infty} \frac {B_{2k+2}} {4k(k+1)z^{2k}},
\end{equation}
where $A$ is a constant and $B_n$ are Bernoulli numbers. Then
\be
  \label{eq:asymptotic-Barnes}
 \big[\log G(1+z) + \log G(1-z)\big] =
  \frac 1 {6} - 2\log A + (\frac {z^2} {2} - \frac {1} {12}) \log (-z^2) - \frac 3 2 z^2 + \dots
\ee
where, two constant terms do not depend on $a$. The $z$ square terms cancel with each other because of the valid condition. We define
\be
F(z)=(\frac {z^2} {2} - \frac {1} {12}) \log (-z^2)\;.
\ee
Under Wick rotation,
\be
F(z_E)=(-\frac {z_E^2} {2} - \frac {1} {12}) \log (z_E^2)\;.
\ee
\subsubsection{Wilson loop in SU(N) linear quiver}
We extend Pestun's formular to quiver with two gauge groups and two bi-fundamentals.\footnote{We found there is a similar extension for Chern-Simons quivers recently by Kapustin etc~\cite{Kapustin:2009kz}, which can give consistent results for Wilson loops in ABJM.} \be\label{loopintegrateQ}
W^{(1)}=Z^{-1}\int \prod_{i< j} (\lambda_i^{(1)}-\lambda_j^{(1)})^2 \, (\prod_i^{N-1} e^{-\frac { 4 \pi^2} {g^{2}} {\lambda_i^{(1)}}^2} d\lambda^{(1)}_i)\frac{ \prod_{i\neq j} H(i\lambda^{(1)}_i-i\lambda^{(1)}_j) }
{ \big[\prod_{i,j} H(i\lambda^{(1)}_i-i\lambda^{(2)}_j)\big] }\sum_i^{N}e^{2\pi \lambda^{(1)}_i},
\ee
where $\lambda^{(2)}$ is from the Lie $G^{(2)}$. $W^{(2)}$ can be defined similarly.
\subsubsection{Saddle point solutions}
Under $N\rightarrow \infty$, from (\ref{loopintegrate}), we obtain the saddle-point equation of motions:
\be
{8\pi^2\over g^2N}\lambda_i+2F'(\lambda_i)-{1\over N}\sum_{j(\neq i)}F'(\lambda_i-\lambda_j)={2\over N}\sum_{j(\neq i)}{1\over \lambda_i-\lambda_j}\;.
\ee
We extend (\ref{loopintegrate}) to quiver gauge theory and in quiver theory composed of two groups $SU(N)$, and bi-fundamental matters we obtain
\be
{8\pi^2\over g_{(1)}^2N}\lambda^{(1)}_i+{1\over N}\sum_jF'(\lambda^{(1)}_i-\lambda^{(2)}_j)-{1\over N}\sum_{j(\neq i)}F'(\lambda^{(1)}_i-\lambda^{(1)}_j)={2\over N}\sum_{j(\neq i)}{1\over \lambda^{(1)}_i-\lambda^{(1)}_j}\;.
\ee

\be
{8\pi^2\over g_{(2)}^2N}\lambda^{(2)}_i+{1\over N}\sum_jF'(\lambda^{(2)}_i-\lambda^{(1)}_j)-{1\over N}\sum_{j(\neq i)}F'(\lambda^{(2)}_i-\lambda^{(2)}_j)={2\over N}\sum_{j(\neq i)}{1\over \lambda^{(2)}_i-\lambda^{(2)}_j}\;.
\ee
Observing the above two equations, we find that by changing variables, the average loop has the same equation of motion as in the $\cN=4$ SYM, with the effective coupling\footnote{We found consistent result was obtained by analyzing an explicit $\cN=2$ example in S.J.Rey's talk in Strings 2009~\cite{Rey:2009talk}. Also we found the related discussion in T.Suyama's talk~\cite{Suyama:talk}.}
\be
 {1\over \tilde\lambda}={1\over 2}({1\over \tilde\lambda_1}+{1\over \tilde\lambda_2}).
\ee
As a simple extension, in the $N$ $SU(N)$ gauge groups linear quiver, we have
\be
{1\over \tilde\lambda}={1\over N}({1\over \tilde\lambda_1}+{1\over \tilde\lambda_2}+\cdots+{1\over \tilde\lambda_N}).
\ee
where $\tilde\lambda=g^2N$ and $\tilde\lambda_i=g_i^2N$ are t'Hooft couplings. Thus we proved (\ref{coupling}). We expect there is a membrane duality for the $W_{T_N}$ loop operator, whose expectation value is given by (\ref{fieldresult}). This membrane should be one M2 brane with one direction compactified on $\Sigma_2$. We will study this in the following section.

\section{Dual Membrane in supergravity}
 From the geometry approximately AdS$_7\times$ S$^4$ at the boundary, if we put the M5 branes on a Riemann surface, then the IR geometry is AdS$_5\times$ S$^4$ fibered on $\Sigma_2$, the metric of which is given by~\cite{Gaiotto:2009gz}
 \bea\label{background}
 \begin{split}
 ds^2_{11}   =&  ( \pi N l_p^3)^{2/3} { W^{1/3} \over 2}
\left\{   4 ds^2_{AdS_5} +   2 \left[ 4 { ( dr^2
+ r^2 d \beta^2 ) \over (1-r^2)^2 } \right]  + 2  d\theta^2  \right.+
\cr &\left. + { 2 \over W} \cos^2 \theta ( d\psi^2 + \sin^2 \psi d\phi^2  ) +
 { 4 \over W}  \sin^2 \theta ( d \chi +  { 2 r^2 d \beta \over (1-r^2) } )^2
\right\}\;,
\cr
&~~~ W  \equiv  (1 + \cos^2 \theta )\;,
\end{split}
 \eea
where $r,\beta$ parametrize two dimensional hyperbolic space, which is quotiented to be a compact Riemann surface by a discrete group $\Gamma$. Angles $\theta,\psi,\phi,\chi$ denote a topologically S$^4$. Note that central charge is given by
\be
c =  { N^3 \over 3  }  {A_{\Sigma } \over 4 \pi }  =
{ N^3 \over 3 } (g-1) ~,~~~~~~~~A_{\Sigma } = 4 \pi ( g-1) ~,~~~~~~~~~g>1,
\ee
which can be checked to be consistent with quivers with the same genus $g$.
\subsection{Expectation value of Wilson loop at strong coupling}

The Wilson loop operator is expected dual to M2 branes with one world volume direction wrapping one cycle in the Riemann surface $\Sigma_2/\Gamma$. For simplicity, we assume M2 brane embedding is:
\be
z=\sigma_1, t=\tau, r=r(\sigma_2), \beta=\beta(\sigma_2),
\ee
where, $z$ is the AdS radial direction, then the action is
\be
S=-T_{M_2}\int d^3\sigma \sqrt{\det g},\quad g_{ab}={\p X^\mu\over\p \sigma^a}{\p X^\nu\over\p \sigma^b}G_{\mu\nu},\quad T_{M_2}={1\over 2\pi^2 l_P^3}.
\ee
Since $\sigma_2$ is compactified, we can obtain the effective string tension in the AdS$_5$
\be
T_s={\int d\sigma_2\sqrt{g_{\sigma_2\sigma_2}}\over 4\pi^2 l_P^3}.
\ee
Then we have $T_s\sim N^{1/3}l_p^{-2}$, if we use the usual AdS/CFT duality relation $\sqrt{\lambda}={R^2\over \alpha'}$, we have effective $\lambda\sim N^2$. The expectation value of strongly coupled Wilson loop turns\footnote{Here, we subtract the divergent term and use the conformal equivalence between a line and a circle on the boundary.}
\be
\langle W\rangle=e^{-S_{M_2}}=e^{-S_{string}}=e^{g[\gamma]N},
\ee
where, the coupling constant depends only on the compactifying circle $\gamma$ of the M2 brane over $\Sigma_2$. The exponential term is same as the result in (\ref{fieldresult}) by differing a normalized factor. One can check the SUSY of embedding M2 brane in background (\ref{background}). This can be followed by~\cite{Drukker:2009tz}.
\section*{Acknowledgement}
The author acknowledges Wei-Shui Xu, Jian-Feng Wu, Yan Liu, Prof.Bin Chen, Prof.Tian-jun Li for helpful discussions.
The author particularly thanks Prof.Miao Li for warmhearted support.


\begin{thebibliography}{20}
\addtolength{\parskip}{-1ex}

\bibitem{Gaiotto:2009gz}
  D.~Gaiotto and J.~Maldacena,
  ``The gravity duals of N=2 superconformal field theories,''
  arXiv:0904.4466 [hep-th].

\bibitem{Maldacena:1997re}
J.~M.~Maldacena,
``The large $N$ limit of superconformal field theories and supergravity,''
Adv.\ Theor.\ Math.\ Phys.\  {\bf 2} (1998) 231
[Int.\ J.\ Theor.\ Phys.\  {\bf 38} (1999) 1113]
[hep-th/9711200].

\bibitem{Rey}
S.~J.~Rey and J.~T.~Yee,
``Macroscopic strings as heavy quarks in large $N$ gauge theory and anti-de
Sitter supergravity,''
Eur.\ Phys.\ J.\ C {\bf 22}, 379 (2001)
[hep-th/9803001].

\bibitem{Maldacena-wl}
J.~M.~Maldacena,
``Wilson loops in large $N$ field theories,''
Phys.\ Rev.\ Lett.\  {\bf 80}, 4859 (1998)
[hep-th/9803002].

\bibitem{Erickson:2000af}
  J.~K.~Erickson, G.~W.~Semenoff and K.~Zarembo,
  ``Wilson loops in $\cN = 4$ supersymmetric Yang-Mills theory,''
  Nucl.\ Phys.\  B {\bf 582} (2000) 155
  [hep-th/0003055].

\bibitem{Drukker:2000rr}
  N.~Drukker and D.~J.~Gross,
  ``An exact prediction of $\cN = 4$ SUSYM theory for string theory,''
  J.\ Math.\ Phys.\  {\bf 42} (2001) 2896
  [hep-th/0010274].

\bibitem{Nekrasov:2002qd}
  N.~A.~Nekrasov,
  Adv.\ Theor.\ Math.\ Phys.\  {\bf 7}, 831 (2004)
  [arXiv:hep-th/0206161].

\bibitem{Pestun:2007rz}
  V.~Pestun,
  ``Localization of gauge theory on a four-sphere and supersymmetric Wilson
  loops,''
  arXiv:0712.2824.

\bibitem{Witten:1997sc}
  E.~Witten,
  ``Solutions of four-dimensional field theories via M-theory,''
  Nucl.\ Phys.\  B {\bf 500}, 3 (1997)
  [arXiv:hep-th/9703166].

\bibitem{Gaiotto:2009we}
  D.~Gaiotto,
  ``${\mathcal{N}}\!=2$ Dualities,''
  arXiv:0904.2715 [hep-th].

\bibitem{Seiberg:1994aj}

  N.~Seiberg and E.~Witten,
  ``Monopoles, Duality and Chiral Symmetry Breaking in ${\mathcal{N}}\!=2$ Supersymmetric   QCD,''
{\slshape   Nucl.\ Phys.\  B }{\bf 431} (1994) 484
  [arXiv:hep-th/9408099].

\bibitem{Minahan:1996fg}

  J.~A.~Minahan and D.~Nemeschansky,
  ``An ${\mathcal{N}}\!=2$ Superconformal Fixed Point with $E_{6}$ Global Symmetry,''
{\slshape   Nucl.\ Phys.\  B }{\bf 482} (1996) 142
  [arXiv:hep-th/9608047].

\bibitem{Argyres:2007cn}

  P.~C.~Argyres and N.~Seiberg,
  ``S-Duality in ${\mathcal{N}}\!=2$ Supersymmetric Gauge Theories,''
{\slshape   JHEP }{\bf 0712} (2007) 088
  [arXiv:0711.0054 [hep-th]].

\bibitem{Tachikawa:2009rb}
  Y.~Tachikawa,
  ``Six-dimensional $D_N$ theory and four-dimensional SO-USp quivers,''
  JHEP {\bf 0907}, 067 (2009)
  [arXiv:0905.4074 [hep-th]].

\bibitem{Benini:2009gi}
  F.~Benini, S.~Benvenuti and Y.~Tachikawa,
  ``Webs of five-branes and N=2 superconformal field theories,''
  JHEP {\bf 0909}, 052 (2009)
  [arXiv:0906.0359 [hep-th]].

\bibitem{Nanopoulos:2009xe}
  D.~Nanopoulos and D.~Xie,
  ``N=2 SU Quiver with USP Ends or SU Ends with Antisymmetric Matter,''
  JHEP {\bf 0908}, 108 (2009)
  [arXiv:0907.1651 [hep-th]].

\bibitem{Alday:2009aq}
  L.~F.~Alday, D.~Gaiotto and Y.~Tachikawa,
 ``Liouville Correlation Functions from Four-dimensional Gauge Theories,''
  arXiv:0906.3219 [hep-th].
\bibitem{NLC}
  N.~Wyllard,
  ``$A_{N-1}$ conformal Toda field theory correlation functions from conformal
  N=2 SU(N) quiver gauge theories,''
  arXiv:0907.2189 [hep-th].

  D.~Gaiotto,
  ``Asymptotically free N=2 theories and irregular conformal blocks,''
  arXiv:0908.0307 [hep-th].

  D.~Nanopoulos and D.~Xie,
  ``On Crossing Symmmetry and Modular Invariance in Conformal Field Theory and
  S Duality in Gauge Theory,''
  arXiv:0908.4409 [hep-th].

  N.~Drukker, J.~Gomis, T.~Okuda and J.~Teschner,
  ``Gauge Theory Loop Operators and Liouville Theory,''
  arXiv:0909.1105 [hep-th].

  L.~F.~Alday, D.~Gaiotto, S.~Gukov, Y.~Tachikawa and H.~Verlinde,
  ``Loop and surface operators in N=2 gauge theory and Liouville modular
  arXiv:0909.0945 [hep-th].

  G.~Bonelli and A.~Tanzini,
  arXiv:0909.4031 [hep-th].

\bibitem{Chen:2008bp}
  B.~Chen and J.~B.~Wu,
  ``Supersymmetric Wilson Loops in N=6 Super Chern-Simons-matter theory,''
  arXiv:0809.2863 [hep-th].

\bibitem{Rey:2008bh}
  S.~J.~Rey, T.~Suyama and S.~Yamaguchi,
  ``Wilson Loops in Superconformal Chern-Simons Theory and Fundamental Strings
  in Anti-de Sitter Supergravity Dual,''
  JHEP {\bf 0903}, 127 (2009)
  [arXiv:0809.3786 [hep-th]].

\bibitem{Drukker:2008zx}
  N.~Drukker, J.~Plefka and D.~Young,
  ``Wilson loops in 3-dimensional N=6 supersymmetric Chern-Simons Theory and
  their string theory duals,''
  JHEP {\bf 0811}, 019 (2008)
  [arXiv:0809.2787 [hep-th]].

\bibitem{Berkovits:1993zz}
  N.~Berkovits,
  ``A Ten-dimensional superYang-Mills action with off-shell supersymmetry,''
  Phys.\ Lett.\  B {\bf 318}, 104 (1993)
  [arXiv:hep-th/9308128].

\bibitem{Drukker:2009tz}
  N.~Drukker, D.~R.~Morrison and T.~Okuda,
  ``Loop operators and S-duality from curves on Riemann surfaces,''
  JHEP {\bf 0909}, 031 (2009)
  [arXiv:0907.2593 [hep-th]].

\bibitem{Kapustin:2009kz}
  A.~Kapustin, B.~Willett and I.~Yaakov,
  ``Exact Results for Wilson Loops in Superconformal Chern-Simons Theories with
  Matter,''
  arXiv:0909.4559 [hep-th].

\bibitem{Rey:2009talk}
S.J.Rey,``http://strings2009.roma2.infn.it/talks/Rey\underline{ }Strings09.PDF''.

\bibitem{Suyama:talk}
T.Suyama,``http://research.kek.jp/group/www-theory/theory\underline{ }center/SAL/slides/
Suyama\underline{ }090417.pdf''
\end{thebibliography}
\end{document}